\documentclass{elsart}
\usepackage{graphics}
\begin{document}
\begin{frontmatter}
\hyphenation{Coul-omb ei-gen-val-ue ei-gen-func-tion Ha-mil-to-ni-an
  trans-ver-sal mo-men-tum re-nor-ma-li-zed mas-ses sym-me-tri-za-tion
  dis-cre-ti-za-tion dia-go-na-li-za-tion in-ter-val pro-ba-bi-li-ty
  ha-dro-nic he-li-ci-ty Yu-ka-wa con-si-de-ra-tions spec-tra
  spec-trum cor-res-pond-ing-ly}
\vspace{-1cm}

\begin{flushright}
hep-ph/0011284 
\end{flushright}
\title{Baryon wave functions in QCD and integrable models}
\author{V.M.~Braun }
\address{Institut f\"ur Theoretische Physik, 
Universit\"at Regensburg, \\D-93040 Regensburg, Germany\\
         Vladimir.Braun@physik.uni-regensburg.de}
\date{19 June 2000}
\begin{abstract}
 A new theoretical framework is proposed for the description of baryon 
light-cone distribution amplitudes, based on the observation that their scale 
dependence to leading logarithmic accuracy is described by a completely 
integrable model. The physical interpretation is that one is able to find 
a new quantum number that distinguishes partonic components in the nucleon 
with different scale dependence.  
\end{abstract}
\end{frontmatter}
\section{Introduction}
\label{introduction}
The notion of baryon distribution amplitudes refers to the valence 
component of the Bethe-Salpeter wave function at small transverse 
separations and is central for the theory of hard exclusive reactions
involving baryons~\cite{earlybaryon}. As usual for a field theory, extraction of the 
asymptotic behavior (here: zero transverse separation) introduces divergences
that can be studied by the renormalization-group (RG) method.
The distribution amplitude $\phi$ thus becomes a function of the three 
quark momentum fractions $x_i$ and the scale that serves as a UV cutoff
in the allowed transverse momenta. Solving the corresponding RG equations 
one is led to the expansion
\begin{eqnarray}
 \phi(x_i,Q^2) &=& 120 x_1 x_2 x_3 \sum_{N,q}\phi_{N,q}(Q_0^2)
 P_{N,q}(x_i)
\left(\frac{\alpha_s(Q^2)}{\alpha_s(Q_0^2)}\right)^{\gamma_{N,q}/b}
\label{expansion}
 \end{eqnarray}
where the summation goes over all multiplicatively renormalizable 
operators built of three quarks and $N$ derivatives. 
The polynomials 
$P_{N,q}(x_i)$ and anomalous dimensions $\gamma_{N,q}$ are obtained by 
the diagonalization of the mixing matrix for the three-quark operators
\begin{eqnarray*}
B_{k_1,k_2,k_3} &=& (D_+^{k_1} q) (D_+^{k_2} q)(D_+^{k_3} q); \quad 
  k_1+k_2+k_3 =N
\end{eqnarray*}
and $\phi_{N,q}(Q_0^2)$ are the corresponding (nonperturbative) matrix 
elements.

As well known, conformal symmetry allows one to resolve the mixing with 
operators containing total derivatives \cite{ER}--\cite{Nyeo} and requires
that the 'eigenfunctions' $P_{N,q}$ corresponding to different values 
of $N$ are mutually orthogonal with the weight function $ 120 x_1 x_2 x_3$ 
that plays the role of the asymptotic wave function. In difference to mesons,
the conformal symmetry is not sufficient, however, to solve the RG equations:
for each $N$ there exist $N+1$ independent operators with the same conformal 
spin which mix with each other. This mixing produces a nontrivial spectrum 
of anomalous dimensions, see Fig.~1, about which little was known until 
recently. 
\begin{figure}
\centerline{\resizebox{0.7\textwidth}{!}{
 \includegraphics{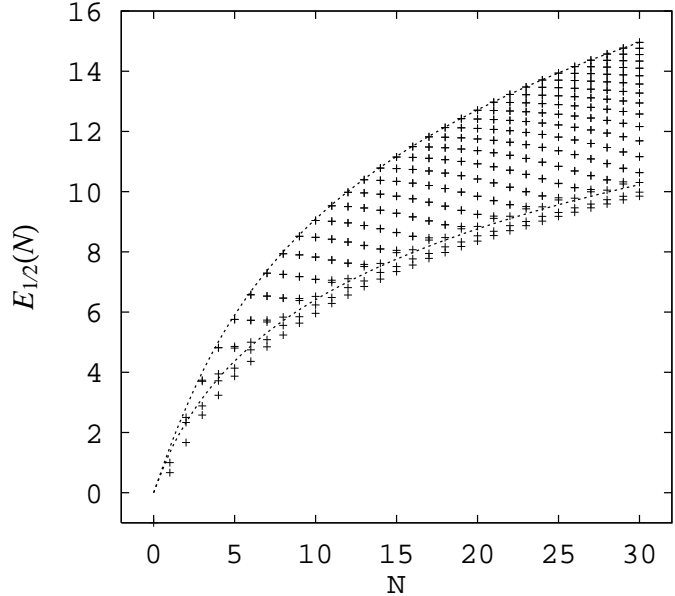}}}
\caption[]{\small The spectrum of anomalous dimensions  
$\gamma_N \equiv (1+1/N_c)E_N+3/2 C_F$ for the
baryon distribution amplitudes with helicity  $\lambda=1/2$. 
The lines of the largest and the smallest
eigenvalues for $\lambda=3/2$ are indicated by  dots
for comparison.}
\label{fig:e12}
\end{figure}
In what follows, I describe a new approach to the scale 
dependence of baryon distribution amplitude developed in \cite{BDM,BDKM}.
The main result is that one is able to identify the second summation 
index $q$ in (\ref{expansion}) with an eigenvalue of a certain conserved
charge. The physical interpretation is that one is able to find a new 
`hidden' quantum number that distinguishes between partonic components 
in the proton with different scale dependence.

\section{Hamiltonian formulation of the evolution equations}
In the usual formulation of the evolution equations the conformal 
invariance is not explicit. It can be made manifest, however, if 
the evolution kernels are rewritten in terms of the generators
of the SL(2) collinear subgroup of conformal transformations.
The scale dependence is different for helicity $\lambda=3/2$ and $\lambda=1/2$
operators and the corresponding evolution kernels can be written in
the following compact form \cite{BDM,BDKM}:
\begin{eqnarray}
 H_{3/2} &=& 2\left(1+\frac{1}{N_c}\right)\sum_{i<k}\Big[
 \psi(J_{ik})-\psi(2)\Big]+\frac32 C_F\,,
\label{3/2}\\
 H_{1/2} &=& H_{3/2}- 2\left(1+\frac{1}{N_c}\right)\left[
    \frac{1}{J_{12}(J_{12}-1)}+
\frac{1}{J_{23}(J_{23}-1)}\right].
\label{1/2}
\end{eqnarray}     
Here $\psi(x)$ is the logarithmic derivative of the $\Gamma$-function
and $J_{ik}, i,k =1,2,3$ are defined in terms of the two-particle
Casimir operators of the $SL(2,R)$ group   
\begin{eqnarray}
  J_{ik}(J_{ik}-1) &=& L_{ik}^2 \equiv (\vec L_i+\vec L_k)^2\,, 
\end{eqnarray}
with $\vec L_i$ being the group generators acting on  the 
i-th quark. In the momentum fraction representation (\ref{expansion})
the generators  take the form \cite{BDKM}
\begin{eqnarray}
 L_{k,0} P(x_i) &=& (x_k\partial_k + 1) P(x_i)\,,\nonumber\\
 L_{k,+} P(x_i) &=& - x_k P(x_i)\,,\nonumber\\
 L_{k,-} P(x_i) &=& (x_k\partial_k^2 + 2\partial_k) P(x_i)\,.
\end{eqnarray}
Solution of the evolution equations corresponds in this language to 
solution of the Schr\"odinger equation
\begin{eqnarray}
  H P_{N,q}(x_i) &=& \gamma_{N,q} P_{N,q}(X_i)
\end{eqnarray} 
with $\gamma_{N,q}$ being the anomalous dimensions.
The $SL(2,R)$  invariance of the evolution equations implies 
that the generators of conformal transformations commute with 
the `Hamiltonians'
\begin{eqnarray}
 [H,L^2]&=& [H,L_\alpha] = 0\,,
\end{eqnarray} 
where $L^2 = (\vec L_1+\vec L_2+\vec L_3)^2$ and 
$L_\alpha = L_{1,\alpha}+L_{3,\alpha}+L_{3,\alpha}$, 
so that the polynomials $P_{N,q}(x_i)$
corresponding to multiplicatively renormalizable operators can be chosen 
simultaneously to be eigenfunctions of $L^2$ and $L_0$:
\begin{eqnarray}
    L^2 P_{N,q} &=& (N+3)(N+2) P_{N,q}\,,\nonumber\\
    L_0 P_{N,q} &=& (N+3) P_{N,q}\,,\nonumber\\
    L_- P_{N,q} &=& 0\,.
\label{conform}   
\end{eqnarray}
The third condition in (\ref{conform}) ensures that the operators do not 
contain overall total derivatives.

Main finding of \cite{BDM} is that the Hamiltonian $H_{3/2}$ possesses an 
additional integral of motion (conserved charge):
\begin{eqnarray}
  Q = \frac{i}{2}[L^2_{12},L^2_{23}] = i(\partial_1\!-\!\partial_2)
(\partial_2\!-\!\partial_3)(\partial_3\!-\!\partial_1) x_1 x_2 x_3
\,,
\quad [H_{3/2},Q] = 0\,. 
\end{eqnarray}
The evolution equation for baryon distribution functions with maximum 
helicity is, therefore, completely integrable. The premium is that 
instead of solving a Schr\"odinger equation with a complicated nonlocal 
Hamiltonian, it is sufficient to solve a much simpler equation
\begin{eqnarray}
    Q P_{N,q}(x_i) = q P_{N,q}(x_i)\,.
\label{Q}
\end{eqnarray}
Once the eigenfunctions are found, the eigenvalues of the Hamiltonian
(anomalous dimensions) are obtained as algebraic functions of $N,q$.

It is necessary to add that the Hamiltonian in (\ref{3/2}) is known as
the Hamiltonian describing the so-called $XXX_{s=-1}$ Heisenberg spin magnet.  
The same Hamiltonian was also encountered in the theory of interacting
reggeons in QCD \cite{FK,Lip}.

\section{Summary of results: Helicity $\lambda=3/2$ distributions}
The equation in (\ref{Q}) cannot be solved exactly, but a wealth of analytic 
results can be obtained by means of the $1/N$ expansion \cite{K,BDKM}.  
%
\begin{figure}
\centerline{\resizebox{0.7\textwidth}{!}{
 \includegraphics{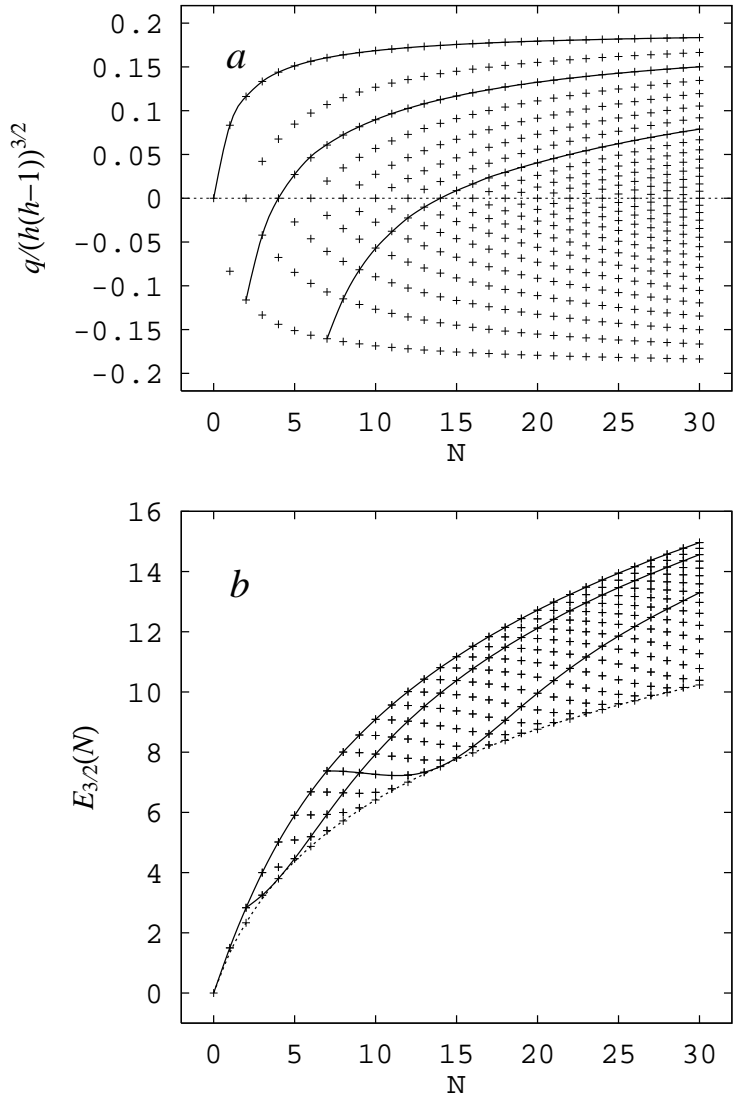}}}
\caption[]{\small The spectrum of eigenvalues for the conserved charge $Q$ (a)
and for the helicity-3/2 Hamiltonian $H_{3/2}$ (b), see text.
Notation: $h=N+3$.}
\label{figure2}
\end{figure}
The general structure of the spectrum is illustrated in Fig.~2. 
It is easy to see that if $q$ is an eigenvalue of $Q$, then $-q$ is 
also an eigenvalue, whereas the Hamiltonian only depends on the absolute 
value $|q|$. It follows that all anomalous dimensions are double degenerate
except for the lowest ones for each {\it even} $N$, corresponding to 
the solution with $q=0$. The corresponding eigenfunctions have a very 
simple form 
\begin{eqnarray}
x_1 x_2 x_3 P_{N,q=0}(x_i) &=& x_1(1-x_1)C^{3/2}_{N+1}(1-2x_1)
+x_2(1-x_2)C^{3/2}_{N+1}(1-2x_2)
\nonumber\\&&{}+x_3(1-x_3)C^{3/2}_{N+1}(1-2x_3)
\label{lowest}
\end{eqnarray}
and the  anomalous dimension is equal to 
\begin{eqnarray}
 \gamma_{N,q=0} &=& \left(1+1/N_c\right)
   \Big[4\psi(N+3)+4\gamma_E-6\Big]
  +3/2 C_F\,.
\label{lowen}
\end{eqnarray}
Furthermore, the eigenvalues of $Q$ lie on trajectories (see Fig.~2) 
corresponding to the semi-classically quantized soliton waves \cite{KK}.
The corresponding trajectories for the anomalous dimensions have a 
rather peculiar form. Each of them can be considered as a separate 
partonic component in the nucleon wave function, in the same spirit as
the leading-twist parton distribution functions in deep-inelastic scattering
arises from the analytic continuation of the anomalous dimensions, giving 
rise to the Altarelli-Parisi splitting function. The asymptotic expansions    
for the charge $q$ and the anomalous dimensions at large $N$ are available 
to the order $1/N^8$ \cite{K,BDKM} and give very accurate results. 
The algebraic structure 
of the spectrum is very complicated. As an example, I present the expression 
for the anomalous dimension $\gamma_{N,q}=(1+1/N_c)E_{N,q}+3/2 C_F$
as a function  of the charge $q$: \cite{K}
\begin{eqnarray}
 E_{N,q} &=& 2 \ln 2 -6 +6\gamma_E + 2\mbox{\rm Re}\sum_{k=1}^3\psi(1+i\eta^3
\delta_k) + { O}(\eta^{-6})\,,
\label{disp}
\end{eqnarray}
where $\eta = \sqrt{(N+3)(N+2)}$ and $\delta_k$, $k=1,2,3$  are the three
roots of the cubic equation
\begin{eqnarray}
 2 \delta_k^3 - \delta_k - q/\eta^3.
\end{eqnarray}
Accuracy of (\ref{disp}) is excellent, as illustrated in Fig.~3.
%
\begin{figure}[t]
\centerline{\resizebox{0.7\textwidth}{!}{
 \includegraphics{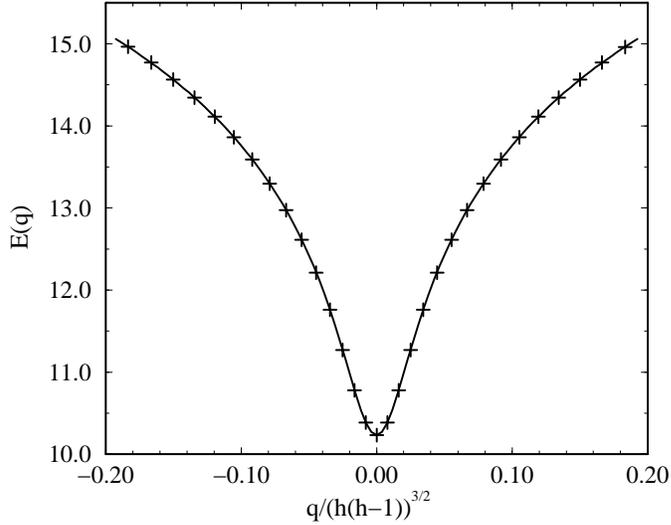}}}
\caption[]{\small The dependence of the energy ${E}$ on the
charge $q$ for $N=30$. The solid curve is calculated
using Eq.~(\protect{\ref{disp}}) and the exact values of energy
for quantized $q$ are shown by crosses.}
\label{figure4}
\end{figure}

\section{Helicity $\lambda=1/2$ distributions}

The additional term in $H_{1/2}$ spoils integrability but can be considered
as a small correction for the most part of the spectrum \cite{BDKM}. 
Its effect on the two lowest levels is drastic, however. To illustrate this,
consider the flow of energy levels for the Hamiltonian 
 $H(\epsilon) = \sum_{i<k}\Big[
 \psi(J_{ik})-\psi(2)\Big] - \epsilon \Big[1/L_{12}^2+1/L_{23}^2\Big]$
(cf. (\ref{3/2}), (\ref{1/2}))
as a function of an auxiliary parameter $\epsilon$, see Fig.~4.
\begin{figure}
\centerline{\resizebox{0.7\textwidth}{!}{
 \includegraphics{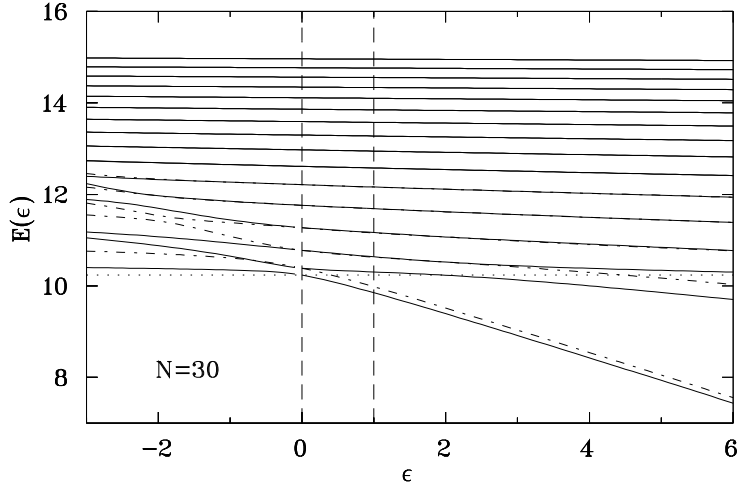}}}
\caption[]{\small The flow of energy eigenvalues for the Hamiltonian
   ${H}(\epsilon)$ for $N=30$.
The solid and the dash-dotted curves show
the parity-even and parity-odd levels, respectively.
The two vertical dashed lines
indicate ${H}_{3/2}\equiv{ H}(\epsilon=0)$ and
${H}_{1/2}\equiv{H}(\epsilon=1)$, respectively (up to the color factors). 
The horizontal dotted line shows position of the `ground state' given by
Eq.~(\protect{\ref{lowen}}).
}
\label{figure7}
\end{figure}
It is seen that the two lowest levels decouple from the rest of the 
spectrum and are separated from it by a finite mass gap. As shown in 
\cite{BDKM}, this phenomenon can be interpreted as binding of the 
two quarks with opposite helicity and forming a scalar ``diquark''.
The effective Hamiltonian for the low-lying levels can be constructed 
and turns out to be a generalization of the famous Kroning-Penney problem
for a particle in a $\delta$-function type periodic potential.
The value of the mass gap between the lowest and the next-to-lowest 
anomalous dimensions at $N\to\infty$ can be calculated combining the small-$\epsilon$
and the large-$\epsilon$ expansions and is equal to 
\begin{eqnarray}
  \Delta\gamma = 0.32\cdot(1+1/N_c)
\end{eqnarray}
in agreement with the direct numerical calculations.

\section{Further developments}
The approach developed in \cite{BDM,BDKM} turns out to be general and 
is applicable to the analysis of all three-parton systems in QCD, 
albeit with some modifications. The extension to three three-gluon
operators is straightforward but tedious, since elementary fields 
have a higher conformal spin \cite{Belitsky1}. Probably more interesting 
from both phenomenological and mathematical point of view 
are the applications to quark-antiquark-gluon systems. 
The corresponding evolution equations are integrable in the limit 
of large number of colors \cite{BDM} and reduce to a different type 
of integrable models - the so-called open chains. A rather detailed 
analysis of integrable quark-antiquark-gluon operators has been given in
\cite{BDM,Belitsky2,DKM} and, most recently,  a concrete phenomenological 
application of this method to the structure function $g_2(x,Q^2)$ was 
considered in work \cite{BKM00}.

To summarize, a powerful mathematical framework has been developed 
for the description of the evolution of baryon distribution amplitudes.
The mathematical structure of evolution equations is very elegant 
and reveals certain qualitative features of the distribution amplitudes.
A lot of analytical results is obtained, in different limits.
The formalism is general and was applied already 
to the other existing three-parton distributions. 
In a more general context,  
the integrability of evolution equations reveals an additional 
symmetry of QCD and its close relation to exactly solvable statistical 
models. Remarkably enough, the same symmetry has been observed 
in the studies of the Regge asymptotics of three-gluon distributions. 
All these features are not seen at the level of QCD Lagrangian and their 
origin has to be understood better.

\section*{Acknowledgments}
The author is grateful
 to S.~Derkachov, G.~Korchemsky and A.~Manashov
for a very rewarding collaboration on the subject of this report. 
Special thanks are due to H.C.~Pauli for the invitation to this conference.

\end{document}